\documentclass[10pt]{article}

\usepackage{amssymb}
\usepackage{amsmath}

\usepackage[margin=1.5in]{geometry}
\usepackage[]{graphicx}
\usepackage[]{cite}
\usepackage[font=small, margin=0.25in]{caption}
\usepackage{subcaption}
\usepackage{siunitx}

\renewcommand{\Re}{\operatorname{Re}}

\begin{document}

\title{Leveraging Continuous Material Averaging for Inverse Electromagnetic Design}
\author{Andrew Michaels and Eli Yablonovitch}
\maketitle

\begin{abstract} 
    Inverse electromagnetic design has emerged as a way of efficiently designing active and passive electromagnetic devices. This maturing strategy involves optimizing the shape or topology of a device in order to improve a figure of merit--a process which is typically performed using some form of steepest descent algorithm. Naturally, this requires that we compute the gradient of a figure of merit which describes device performance, potentially with respect to many design variables. In this paper, we introduce a new strategy based on smoothing abrupt material interfaces which enables us to efficiently compute these gradients with high accuracy irrespective of the resolution of the underlying simulation. This has advantages over previous approaches to shape and topology optimization in nanophotonics which are either prone to gradient errors or place important constraints on the shape of the device. As a demonstration of this new strategy, we optimize a non-adiabatic waveguide taper between a narrow and wide waveguide. This optimization leads to a non-intuitive design with a very low insertion loss of only \SI{0.041}{\decibel} at \SI{1550}{\nano\meter}. 
\end{abstract}

\section{Introduction}

As integrated photonics continues to mature, we have witnessed a growing desire for more compact and efficient passive optical components.  At the same time, our ability to satisfy these demands using either analytic methods or by tuning only a small set of device parameters is becoming increasingly difficult.  To combat this trend, more sophisticated optimization methods are proving very useful. In particular, topology and shape optimization enable us to efficiently design devices with hundreds, thousands, or even millions of independent design parameters.  In the case of shape optimization, the boundaries between different materials composing an initial structure are modified in order to minimize a desired figure of merit (i.e. a function which quantifies the performance of the device being optimized)\cite{haftka_structural_1986}.  Similarly, in the case of topology optimization, a starting structure is modified in order to minimize a figure of merit, however fewer constraints are placed on how the structure can evolve and in particular its topology may be modified (e.g. the creation or elimination of holes)\cite{bendsoe_generating_1988}. 

Shape and topology optimization have found extensive application in structural mechanics, and work on both methods dates back more than 30 years \cite{haftka_structural_1986,bendsoe_generating_1988,sigmund_design_1997,allaire_structural_2004}.  Despite the maturity of this field, comparatively limited application of these methods has found its way into the optics and photonics community.  Early work on gradient-based optimization of microwave devices \cite{hong_systematic_1997,chung_optimal_2000} demonstrated application of shape optimization to electromagnetics.  This work was followed by the application of topology optimization to photonic crystals \cite{borel_topology_2004,veronis_method_2004} and later to a greater variety of passive photonic components\cite{jensen_topology_2011}.  Within the last five years, however, the photonics community has witnessed a steady rise in interest in these optimization techniques as demonstrated by the numerous optimizations of efficient splitters, couplers, etc\cite{lu_nanophotonic_2013,keraly_adjoint_2013,frandsen_topology_2014,piggott_inverse_2015, sigmund_nanostructured_2016,frellsen_topology_2016,piggott_fabrication-constrained_2017,lin_topology_2017,wang_adjoint_2018}.

A large portion of this prior work has focused on implementing topology optimization techniques and has emphasized problems in which the design space consists of thousands or even millions of parameters. In particular, the idea of  choosing the permittivity at each point in a discretized domain as independent design variables has gained popularity.  While this approach has proven useful for generating solutions without requiring significant intuition about the appearance of the final structure, it suffers two primary disadvantages. First, in order to ensure that the device can be fabricated, a final post-processing step is often required to convert a grayscale material distribution to a binary material distribution \cite{elesin_design_2012}.  As a result, the final solution may not be truly optimal or may require additional shape optimization steps. Second, there are many problems in which either the geometry of the structure is constrained in some way or for which topological changes are unnecessary (for example, a one dimensional grating coupler consisting of strictly rectangular segments).  In these cases, topology optimization techniques may be unnecessary or even inappropriate.

Shape optimization serves as a remedy to both of these problems: it can be used in order to fine-tune the result of a topology optimization and is also well suited to handling problems in which the general shape of the structure is known beforehand. Shape optimization has an additional benefit in that it gives us considerable freedom over the choice of design space--we are free to choose arbitrary parameters, such as the length of a rectangle or the position of a circle, as independent design parameters.

In previous work, shape optimization has frequently been used in conjunction with simulation methods that use an unstructured mesh (notably either a finite element method\cite{haftka_structural_1986} or a variant of the finite difference time domain technique\cite{chung_optimal_2000}).  In the nanophotonics community, however, the finite difference time domain method (FDTD) is preferred due to its relative simplicity and computational efficiency.  Unlike the finite element method, FDTD represents materials on a rectangular grid which is not conducive to the representation of non-rectangular material boundaries.  This poses an even greater difficulty to shape optimization as the rectangular grid makes it difficult to represent continuous modifications to material boundaries, a process which is essential to computing accurate gradients of a figure of merit (or sensitivity analysis).  These gradients are helpful to the efficient minimization of the figure of merit, and it is thus desirable that we devise a way to achieve continuous perturbations to boundaries expressed on a rectangular grid.

\begin{figure}[htp!]
    \begin{center}
        \includegraphics[width=\textwidth]{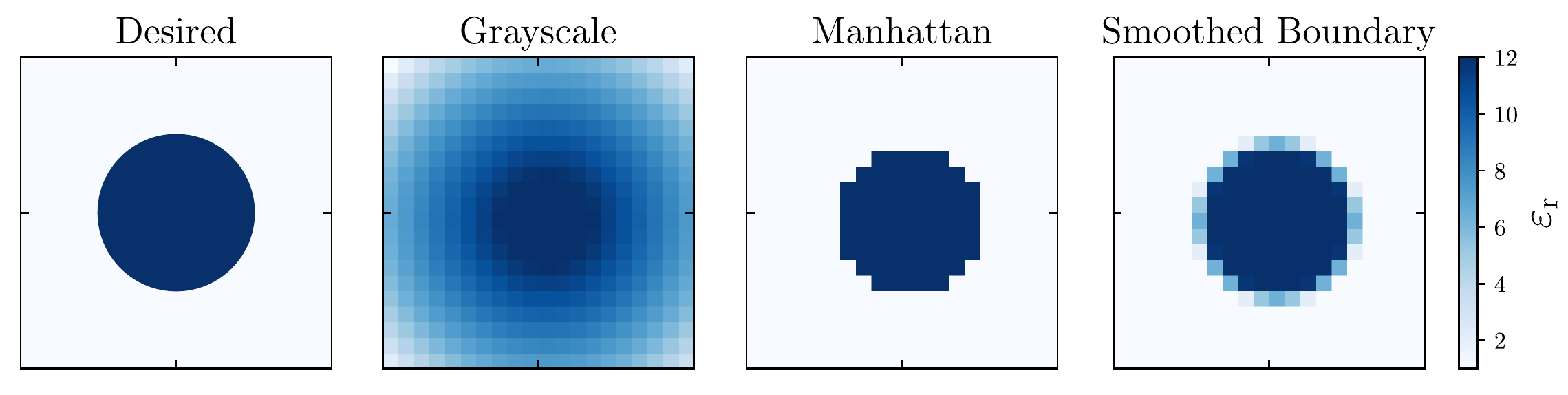}
    \end{center}
    \caption{Comparison of different strategies for representing abrupt material boundaries on a rectangular grid. On the far left, the desired structure is shown (a circle). The next image depicts a grayscale permittivity distribution which yields an approximation of the circle when a threshold is applied. The third image depicts a strictly binary grid (``Manhattan'') representation of the circle. The final picture depicts the circle represented using our own grid smoothing method which best captures the circular boundary.}
    \label{fig:grid_comparison}
\end{figure}

To this end, we have implemented \emph{continuous} boundary smoothing.  Our boundary smoothing is effectively a hybrid of grayscale and manhattan representations of material distributions as depicted in Fig. \ref{fig:grid_comparison}. Unlike the latter two methods, our boundary smoothing allows the material value to change smoothly between two values only in grid cells which intersect the boundary as shown in Fig. \ref{fig:grid_comparison}. Infinitesimal modifications to the boundary of a material are reflected in infinitesimal changes to the effective permittivity of the intersecting grid cells.  This enables us to calculate accurate gradients of a figure of merit with respect to many arbitrary variables with only two simulations using the adjoint method.  These gradients can then be fed into a variety of minimization algorithms in order to optimize our structure.  This intuitive boundary smoothing method, which in the past has been considered primarily for the purpose of improving simulation accuracy, can enable us to easily optimize electromagnetic structures which would otherwise prove very challenging.

In this paper, we will briefly review the adjoint method which is used to compute the gradient of the figure of merit and highlight the importance of smooth boundary modifications.  We will then present and validate our boundary smoothing method and finally use it to optimize an efficient short waveguide taper with a constrained feature size.

\section{The Adjoint Method}

Optimization (or ``inverse design'') of an electromagnetic device involves modifying an initial design in order to maximize (or minimize) a figure of merit which describes the device's performance.  In order to modify the design, we specify design parameters which define the shapes of material boundaries and hence the structure of the device. There are many strategies for choosing the values of our design parameters in order to maximize our figure of merit; of particular interest are gradient-based minimization techniques which are very efficient assuming we can inexpensively compute the gradients of our figure of merit (i.e. the derivative of the figure of merit with respect to each design parameter).  The simplest way to do this is by a brute force approach in which each design parameter is independently varied and a separate simulation is run in order to determine how the figure of merit changes.  If $M$ design variables specify the shape of the device, then this method requires $M+1$ simulations per gradient calculation, which quickly becomes impractical for even modest values of $M$.

It turns out that through clever application of the chain rule, we can find the gradient of our figure of merit using only two simulations.  This process is called the \emph{adjoint method} or the \emph{adjoint variable method} and has found widespread application in many fields, including electromagnetics. For the sake of clarity and notation, we will rederive the general expressions of the adjoint variable method and point out some key results relevant to grid smoothing.

Our goal is to minimize a figure of merit $F(\mathbf{E}, \mathbf{H})$ which
depends on the spatially-dependent electric ($\mathbf{E}$) and magnetic ($\mathbf{H}$) fields.  The fields are found by solving Maxwell's equations, given by

\begin{align}
		\mathbf{\nabla}\times \mathbf{E} - i\omega\mu\mathbf{H} &= \mathbf{M} \\
		\mathbf{\nabla}\times \mathbf{H} + i\omega\varepsilon\mathbf{E} &= \mathbf{J}
		\label{eq:maxwells_equations}
\end{align}

\noindent which we have chosen to write in their time harmonic form.  As is the case with most linear partial differential equations, we can discretize these equations on a rectangular grid and rewrite them in matrix form as

\begin{equation}
		A \vec x = \vec b
		\label{eq:Ax_b}
\end{equation}

\noindent where $A$ is a matrix containing the curls expressed using finite
differences and the permittivity and permeability at each point in the
discretized space, $\vec x$ is a vector containing the electric and magnetic
fields at all  points in space, and $\vec b$ contains the inhomogenous electric and magnetic current density at all points in space. Depending on how unknowns are ordered when assembling the system, $A$, $\vec x$, and $\vec b$ might take the form

\begin{equation}
    A = \left(\begin{array}{cc}
        i\omega\varepsilon(\mathbf{r}) & \nabla\times\\
        \nabla\times & -i\omega\mu(\mathbf{r})
        \end{array}\right) 
        , \;
    \vec{x} = \left(\begin{array}{c}
        \vec{E}\\
        \vec{H}
        \end{array}\right)
    , \;
    \vec{b} = \left(\begin{array}{c}
        \vec{J}\\
        \vec{M}
        \end{array}\right)
    \label{eq:matrix_and_vectors}
\end{equation}

\noindent where $\vec{E}$, $\vec{H}$, $\vec{J}$, and $\vec{M}$ are $N\times 1$ vectors containing the discretized field and current density values. In our discretized world, our figure of merit becomes a function of $\vec x = (\vec E \;\vec H)^T$ which we write as $F(\vec x)$.

If we had direct control over the electric and magnetic fields in $\vec x$, finding the gradient of $F(\vec x)$ would be require only simple differentiation with respec to $\vec E$ and $\vec H$. Instead of directly controlling $\vec E$ and $\vec H$, we have control over the permittivity and permeability defined everywhere in space which may be specified using a structured set of design parameters (like the dimensions of shapes, the positions of shapes, the coordinates of polygon vertices, etc).  If we write this set of design variables as $\vec{p} = \left\{p_1, p_2, \cdots, p_M\right\}$, then the gradient we are interested in is the set of derivatives of $F$ with respect to each $p_i$, i.e. 

\begin{equation}
		\vec{\nabla}_p F = \left[\frac{\partial F}{\partial p_1}, \frac{\partial F}{\partial p_2}, \cdots,
		\frac{\partial F}{\partial p_M}\right]
		\label{eq:gradient_F}
\end{equation}

To find these derivatives, we begin by applying chain rule when differentiating $F(\vec{x})$.  Consider first the $i$'th derivative for the simple case in which $F$ is an explicit function of the electric and magnetic fields only:

\begin{equation}
		\frac{\partial F}{\partial p_i} = 2\Re\left\{ \frac{\partial F}{\partial \vec x} \frac{\partial \vec x}{\partial p_i} \right\}
		\label{eq:dFdpi_chainrule}
\end{equation}

\noindent Because the fields in $\vec{x}$ are complex valued, we must be careful when taking their derivatives, hence the appearance of the $2\Re\{\cdots\}$.

Notice that the derivative $\partial F / \partial \vec x$ is already known since the figure of merit is an explicit function of the electric and magnetic fields.  The second term in (\ref{eq:dFdpi_chainrule}), $\partial \vec x/\partial p_i$, remains to be found. This is accomplished by directly differentiating our system of equations given in Equation (\ref{eq:Ax_b}) with respect to $p_i$ and multiplying by $A^{-1}$, which yields

\begin{align}
        \frac{\partial}{\partial p_i} \left(A \vec x\right) &= \frac{\partial \vec b}{\partial p_i} \nonumber\\
        \Rightarrow A\frac{\partial \vec x}{\partial p_i} &= \frac{\vec b}{\partial p_i} - \frac{\partial A}{\partial p_i}\vec x \nonumber \\
		\Rightarrow \;\; \frac{\partial \vec x}{\partial p_i} &= A^{-1}\left(\frac{\partial \vec b}{\partial p_i} - \frac{\partial A}{\partial p_i}
		\vec x\right) \;\; .
		\label{eq:dxdp}
\end{align}

\noindent Notice that the Maxwell operator $A$ contains the distribution of permittivity and permeability in the system which is directly controlled by the design parameters $\vec p$. Therefore, $\partial A / \partial p_i$ is known or assumed to at least be easily computable.  The derivative of the current sources with respect to the design parameters, $\partial \vec b / \partial p_i$, can be calculated, although in most cases we will assume that the inputs to the system are fixed and this term will be zero. In this case, Equation (\ref{eq:dxdp}) becomes

\begin{equation}
    \frac{\partial \vec x}{\partial p_i} = -A^{-1}\frac{\partial A}{\partial p_i} \vec x \;\; .
    \label{eq:dxdp_nosrc}
\end{equation}

    Substituting this expression for $\partial \vec x / \partial p_i$ in Equation (\ref{eq:dFdpi_chainrule}), we find an expression for the $i$'th derivative of $F$ in terms of known quantities:

\begin{equation} 
    \frac{\partial F}{\partial p_i} = -2\Re\left\{ \frac{\partial F}{\partial \vec x} A^{-1}\frac{\partial A}{\partial p_i} \vec x\right\} \label{eq:dFdpi_almost_there}
\end{equation}

\noindent This expression can be written in a more enlightening way by introducing a new vector given by

\begin{equation} 
    \vec y^T = \frac{\partial F}{\partial \vec x} A^{-1} 
    \label{eq:yT} 
\end{equation}

\noindent which we can rewrite in the form $A \vec x = \vec b$ by multiplying by $A$ and taking the transpose of both sides:

\begin{equation}
     A^T \vec y = \left(\frac{\partial F}{\partial \vec x}\right)^T
    \label{eq:ATy_equals_b}
\end{equation}

\noindent Solving this expression for $y$ is \emph{similar} to solving Maxwell's equations where $\partial F / \partial \vec x$ acts as the current sources. In general, the discretized form of Maxwell's equations is not symmetric, and therefore the forward and adjoint equations are not identical. Substituting Equation (\ref{eq:yT}) into (\ref{eq:dFdpi_almost_there}), we obtain a final expression for the derivative of our function $F$ with respect to the design variables of the system:

\begin{equation} 
    \frac{\partial F}{\partial p_i} = -2\Re\left\{ \vec y^T \frac{\partial A}{\partial p_i} \vec x \right\} \; .
    \label{eq:dFdpi_final_simple} 
\end{equation}

In order to solve for all $M$ derivatives of $F$ with respect to $p_i$, we need to compute the physical electric and magnetic fields represented by $\vec x$ as well as a second set of non-physical ``adjoint'' fields represented by $\vec y$. We can intuitively think about these adjoint fields as the fields that are produced by injecting the desired output fields (which arise from currents $\partial F/\partial \vec x$) into the system and running the whole system backwards. Solving for $\vec x$ and $\vec y$ each correspond to a single ``forward'' and ``adjoint'' simulation, respectively, and thus solving for the gradient of $F$ requires two simulations, independent of the number of design variables.  This is the great advantage of the adjoint method. 

In addition to the forward and adjoint simulations, an essential component of the adjoint method is the accurate calculation of $\partial A / \partial p_i$ which contains the information about how the distribution of materials in the system is controlled by the design parameters $\vec p$. In particular, the discretized equations which are assembled into $A$ can be ordered such that the permittivity and permeability values are contained in the diagonal of $A$ as indicated by Equation (\ref{eq:matrix_and_vectors}). Typically, when working on a rectangular grid, the discretization of the problem will remain unchanged as the design variables are modified. If this is true and the materials present are either isotropic or diagonally anisotropic, then all of the off-diagonal elements of the system matrix will not change with respect to changes to the design variables and hence the off-diagonal elements of $d A/d p_i$ will be zero. This leaves only the diagonal elements which contain the derivatives of the permittivity and permeability at each point in space. In this case, assuming all permittivities and permeabilities are isotropic, the derivatives of the figure of merit are greatly simplified to


\begin{equation}
    \frac{\partial F}{\partial p_i} = 2\omega\;\text{Im}\left\{\sum\limits_j \frac{\partial \varepsilon_j}{\partial p_i} \mathbf{E}_j \cdot \mathbf{E}_j^{\text{adj}} - \sum\limits_j \frac{\partial \mu_j}{\partial p_i} \mathbf{H}_j \cdot \mathbf{H}_j^{\text{adj}}\right\}
    \label{eq:dFdpi_most_simplified}
\end{equation}

\noindent where the field quantities with the superscript ``adj" are contained in $\vec y$. It is interesting to note that this result is consistent with derivations of the continuous adjoint method\cite{keraly_adjoint_2013} with the exception that Maxwell's equations are not assumed to be symmetric in our ``discrete'' formulation. Based on Equation (\ref{eq:dFdpi_most_simplified}), it is apparent that our ability to accurately compute the gradient of our figure of merit is contingent on our ability to form a continuous relationship between the design parameters of the system and the distribution of permittivity and permeability (were this not the case, $\partial \epsilon / \partial p_i$ and $\partial \mu / \partial p_i$ would be ill-defined). In the case of representing material boundaries on a rectangular grid, this is not straightforward.

To overcome this issue, we use grid smoothing techniques which allow us to project continuously defined boundaries onto a rectangular grid.  If a boundary intersects a grid cell, an intermediate material  value between the  two values on either side of the boundary is assigned to the intersected cell.  This allows us to make very small perturbations to the continuously-defined boundaries (i.e. boundary shifts smaller than the width of a grid cell) which can be modeled as a slight change in the permittivity/permeability in the grid cells which intersect the boundary. Because these changes to permittivity/permeability can be made in a smooth and continuous way, we can calculate $\partial \epsilon / \partial p_i$ and $\partial \mu / \partial p_i$ accurately even on a rectangular grid.  In the next section, we explain this grid smoothing process in detail.

\section{Grid Smoothing}

When trying to represent an abrupt interface between two materials on a rectangular grid, we inevitably run into the  problem of staircasing: representing curved or diagonal boundaries between two different materials on a rectangular grid results in a jagged interface which conforms to the underlying grid. In addition to compromising simulation accuracy, the rectangular nature of the grid poses significant challenges to calculating sensitivities since perturbations smaller than a grid cell are not possible.

A considerable amount of work has been done to improve the treatment of non-rectangular boundaries with finite difference methods for the purpose of improving simulation accuracy. In particular, modification of the FDTD equations have been successfully employed in order to improve the simulation accuracy \cite{jurgens_finite-difference_1992,dey_modified_1998} and the introduction of effective permittivity at material interfaces \cite{dey_conformal_1999,kaneda_fdtd_1997,farjadpour_improving_2006,oskooi_accurate_2009} has been demonstrated to improve simulation accuracy in many situations.  This process of computing effective intermediate material values, which we refer to as ``grid smoothing,'' is particularly relevant to shape optimization as it provides us with a way to achieve small perturbations to the material boundaries represented on a rectangular grid.

\begin{figure}[htp!]
    \centering
    \includegraphics[width=0.8\textwidth]{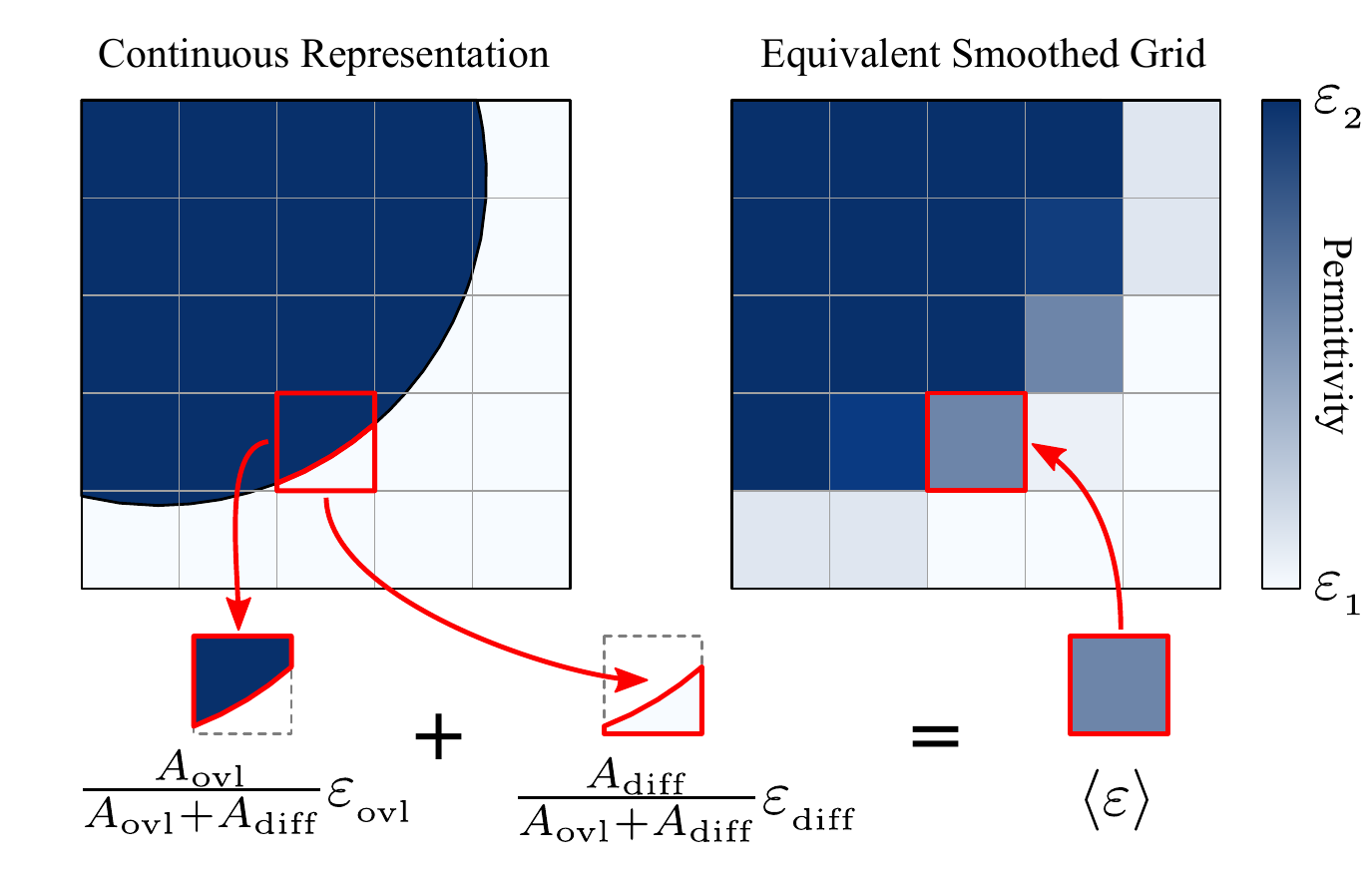}
    \caption{Visual depiction of grid smoothing process. Internally, all material boundaries of the system are represented using polygons which are defined in a continuous domain. These shapes are then mapped onto a rectangular grid by computing the average value of permittivities and permeabilities which overlap with each cell in the grid. This mapping is achieved by computing the overlap area between grid cells and material domains.}
    \label{fig:grid_smoothing_explanation}
\end{figure}

In order to demonstrate this, we have implemented a simple form of grid smoothing using weighted averages which is depicted in Fig. \ref{fig:grid_smoothing_explanation}.  In a given grid cell, the effective permittivity in a 2D domain is given by

\begin{equation}
		\langle\varepsilon(i,j)\rangle = \frac{1}{\Delta x \Delta y}\sum\limits_{k} C_k \varepsilon_{k}(i,j)
		\label{eq:grid_smoothing}
\end{equation}

\noindent where $C_k$ is the overlap area between the k'th material domain and the grid cell at location $i,j$ and $\Delta x$ and $\Delta y$ are the grid cell width and height, respectively.  For the purpose of computing derivatives, it is essential that $C_k$ be computable with high precision.  We accomplish this by representing all boundaries in the system as piecewise linear functions (i.e. polygons) and then computing intersections between the material domains and the grid-cells that intersect the boundaries of those domains. Because these piecewise linear functions are stored with very high (or even arbitrary) numerical precision, infinitesimal modifications to the boundaries are reflected by infinitesimal modifications to the local effective permittivity and permeabilities on the grid. This process is contingent on our ability to efficiently find polygon intersections. Fortunately, this has long been a topic of great importance in computational geometry \cite{preparata_computational_1985}. Due to the maturity of this field, efficient algorithms for finding the intersection between polygons are readily available, making this simple form of grid smoothing relatively straightforward to implement.

\begin{figure}[htp!] \centering \begin{subfigure}[b]{0.425\textwidth}
				\includegraphics[width=\textwidth]{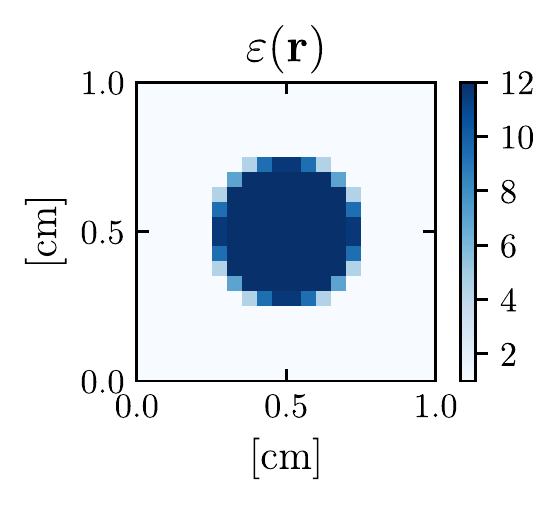} \caption{} \end{subfigure} \hskip -1em
		\begin{subfigure}[b]{0.425\textwidth}
				\includegraphics[width=\textwidth]{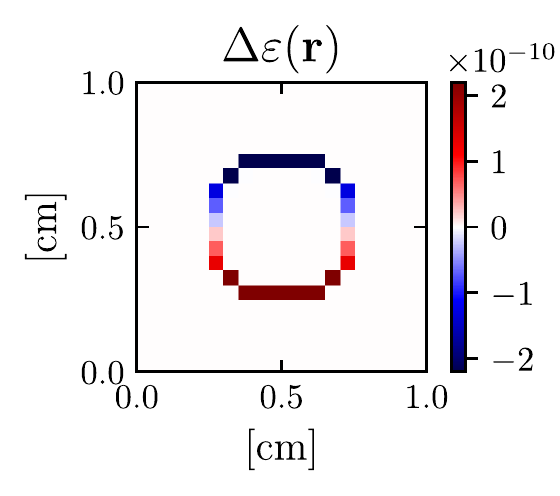} \caption{} \end{subfigure}
		\caption{Demonstration of grid smoothing for a 0.5 cm diameter dielectric circle.  (a) shows the 
				smoothed grid computed for the circle on an intentionally coarse grid in order to clearly show
				the averaging which occurs at the circle's boundary. (b) shows the difference in permittivity between
				the grid shown in (a) and the grid corresponding to the same circle which has been shifted in the y
				direction by $10^{-12}$ cm.  The difference in permittivity is correspondingly small, highlighting the
		continuous nature of our grid smoothing.} 
		\label{fig:grid_smoothing}
\end{figure}

A demonstration of this process are depicted in Fig. \ref{fig:grid_smoothing}.  In (a), the smoothed permittivity for a 0.5 cm diameter circle is shown on an intentionally coarse grid.  As a result of the smoothing process, the grid cells at the boundary of the circle are filled with an effective permittivity whose value is between the permittivity inside of the circle and the permittivity surrounding the circle.  We then test the continuous nature of this smoothing by displacing the circle in the y direction by $10^{-12}$ cm.  The change in the permittivity as a result of this displacement is shown in (b).  The small size of the displacement is reflected by the correspondingly small change in permittivity in the grid cells at the circle's outer boundary.

With continuous grid smoothing at our disposal, the derivatives $\partial A / \partial p_i$ are easily computed using finite differences.  One at a time, each design variable of the system $p_i$ is perturbed by a small amount (e.g. $10^{-8} \times \Delta x$) and the diagonals of a new matrix $A(p_i+\Delta p)$ is computed.  The derivative is then given approximately by 

\begin{equation}
		\frac{\partial A}{\partial p_i} \approx \frac{A(p_i+\Delta p) - A(p_i)}{\Delta p}
		\label{eq:dAdpi}
\end{equation}

\noindent  which is accurate so long as $\Delta p$ is sufficiently small. It is important to note that this process can be used effectively regardless of how coarse or fine the spatial discretization of the underlying simulation is.  Furthermore, updating $A$ incurs significantly less computational overhead compared to running a new simulation. 

Calculation of this derivative combined with solutions to the forward and adjoint problems provide us with the tools we need to optimize passive integrated photonic devices.  To summarize, the process for optimizing a structure using grid smoothing and the adjoint method is as follows: 

\par\vspace*{1em}\noindent\makebox[\textwidth][c]{
\begin{minipage}[]{0.7\textwidth}
\begin{enumerate}
        \item Choose initial structure, figure of merit $F(\vec{x})$, and design parameters $\vec{p}$
		\item Calculate figure of merit and its gradient for current structure
			\begin{enumerate}
				\item Run forward simulation (i.e. solve $Ax=b$)
                \item Calculate $F(\vec{x})$ and $\partial F / \partial \vec{x}$
                \item Run adjoint simulation according to Eq. \ref{eq:ATy_equals_b} to find $y^T$
                \item Perturb design variables one at a time in order calculate $dA/dp_i$ using Eq. \ref{eq:grid_smoothing} and Eq. \ref{eq:dAdpi}
                \item Calculate gradient $dF/d\vec{p}$ using Eq. \ref{eq:dFdpi_final_simple}
			\end{enumerate}
		\item Repeat step 2. as required by minimization algorithm until figure of merit has converged to minimum value
\end{enumerate}

\end{minipage}
}

\vspace*{1em}

\section{Optimization of a Non-adiabatic Waveguide Taper}

\begin{figure}[htp!] 
		\centering 
		\includegraphics[width=0.9\textwidth]{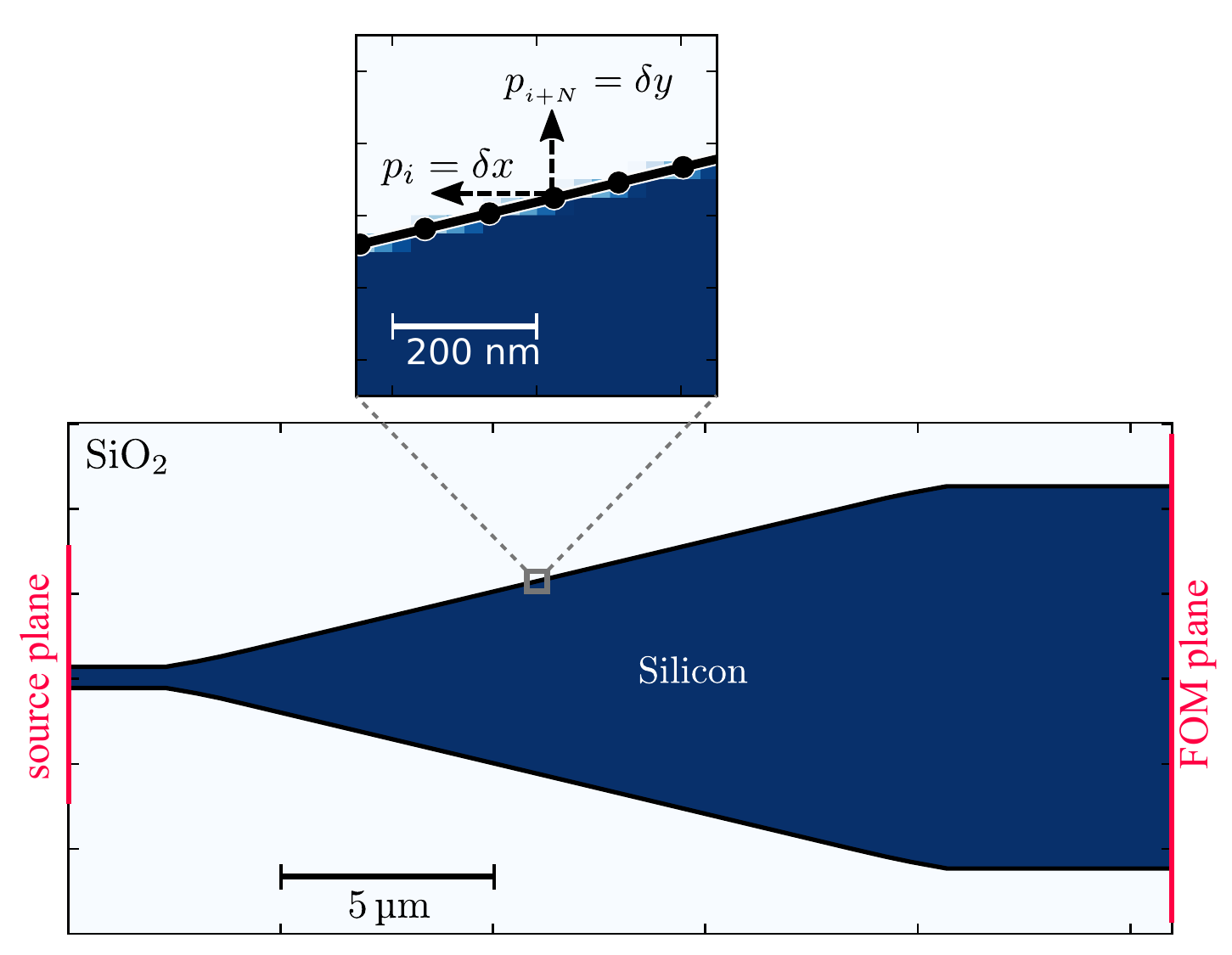}
		\caption{A short \SI{18}{\micro\meter} taper from a \SI{500}{\nano\meter} silicon waveguide to a \SI{9}{\micro\meter} wide silicon waveguide.  The taper is defined as a single polygon containing 200 vertices along its top and bottom diagonal edges.  We choose the displacement of the $x$ and $y$ coordinates of these points (labeled $\delta x$ and $\delta y$) as the design parameters of the system. This taper is used to validate the accuracy of gradients computed using grid smoothing in conjunction with the adjoint method and serves as the starting point of an optimization to demonstrate application of these methods.}
		\label{fig:initial_design} 
\end{figure}

In order to demonstrate the shape optimization process, we have optimized a non-adiabatic waveguide taper which is useful as a compact spot size converter for butt coupling to a fiber or short transition from a waveguide to a grating coupler.  Unlike previous work on the design of short tapers \cite{luyssaert_efficient_2005}, we allow the geometry of the taper to evolve with greater freedom.  This allows us to achieve much more compact transitions than an analytic taper allows \cite{fu_efficient_2014} without sacrificing performance over a large bandwidth.  The taper we optimize is \SI{18}{\micro\meter} long and connects a \SI{500}{\nano\meter} wide input waveguide to a \SI{9}{\micro\meter} wide output waveguide.  The structure is made of \SI{220}{\nano\meter} thick silicon clad in silicon dioxide and is reduced to two dimensions using the effective index method.  The structure is excited with the fundamental TM mode of the \SI{500}{\nano\meter} wide input waveguide using a wavelength of \SI{1550}{\nano\meter}.  We choose the linear taper depicted in Fig. \ref{fig:initial_design} (a) as the initial design.  The taper itself is represented using a polygon with 200 vertices on its top edge and the structure is mirrored about the $x$ axis using symmetry boundary conditions.  We select the design variables of the problem to be the relative $x$ and $y$ displacement of these 200 vertices from their starting positions as depicted in Fig. \ref{fig:initial_design} (b).  We thus use 400 design variables in total.

For all forward and adjoint simulations, we use our own finite difference frequency domain (FDFD) solver. This provides easy access to the internals of the simulator and makes implementing the adjoint simulation straightforward. Although for these examples we use FDFD, our grid smoothing methods are equally applicable to FDTD (a topic which we hope to cover in a future publication). Finally, for all of our simulations, we use a grid resolution of \SI{25}{\nano\meter} in order to achieve sufficient accuracy.

Our goal in optimizing this taper is to maximize the fraction of input power that is coupled into the fundamental TM mode of the wider output waveguide. The figure of merit which quantifies this is the mode matching efficiency integral.  Assuming there is no reflected wave at the output, the continuous form of this mode matching integral, which we derive in Appendix \ref{sec:appendix_mode_match}, is given by

\begin{equation} 
		\eta =  \frac{1}{4 P_m P_\mathrm{src}}\left|\iint\limits_A d\mathbf{A} \cdot \mathbf{E} \times \mathbf{H}_m^*\right|^2
		\label{eq:mode_match_simple} 
\end{equation}

\noindent where $\mathbf{E}$ is the incident electric field, $\mathbf{H}_m$ is the desired magnetic field, $P_\mathrm{src}$ is the source power, and $P_m$ is the power in the desired fields.  The integral is taken over a plane which encompasses the entire desired field profile.  In the case of the waveguide taper, $\mathbf{H}_m$ corresponds to the magnetic fields of the fundamental TM mode of the larger output waveguide while $\mathbf{E}$ is the actual simulated electric field taken along the red line on the right hand side of Fig. \ref{fig:initial_design} (a) (labeled ``FOM plane'').

In this optimization, we not only seek a design that maximizes efficiency, but also a design that can be fabricated.  We accomplish this by applying radius of curvature constraints which prevent the formation of exceedingly small features.  We do this by penalizing the mode match efficiency with a penalty function which reduces the figure of merit when the approximate radius of curvature at each point in the polygon falls below a specified minimum radius of curvature.  The full figure of merit we use is given by

\begin{equation}
    F(\mathbf{E}, \mathbf{H}, \vec{p}) = \eta\left(\mathbf{E}, \mathbf{H}\right) - f_\text{\tiny ROC}(\vec{p})
		\label{eq:figure_of_merit} 
\end{equation}

\noindent where, $\eta$ is the mode match given in Equation (\ref{eq:mode_match_simple}) and $f_\text{\tiny ROC}(\vec{p})$ is a differentiable function of the design variables that is positive when the effective radius of curvature calculated at each vertex drops below a minimum radius of curvature and drops quickly to zero as the calculated radius of curvature increases above this minimum.  In this example, we choose a minimum radius of curvature of \SI{150}{\nano\meter} since it can be easily fabricated.  Because our shape optimization gives us full freedom over the parameterization, the incorporation of such constraints is a straightforward matter of modifying the figure of merit, and, unlike other topology optimization methods \cite{piggott_fabrication-constrained_2017}, requires no additional modification of the underlying minimization process.

\begin{figure}[h] 
		\centering
		\includegraphics[width=0.8\textwidth]{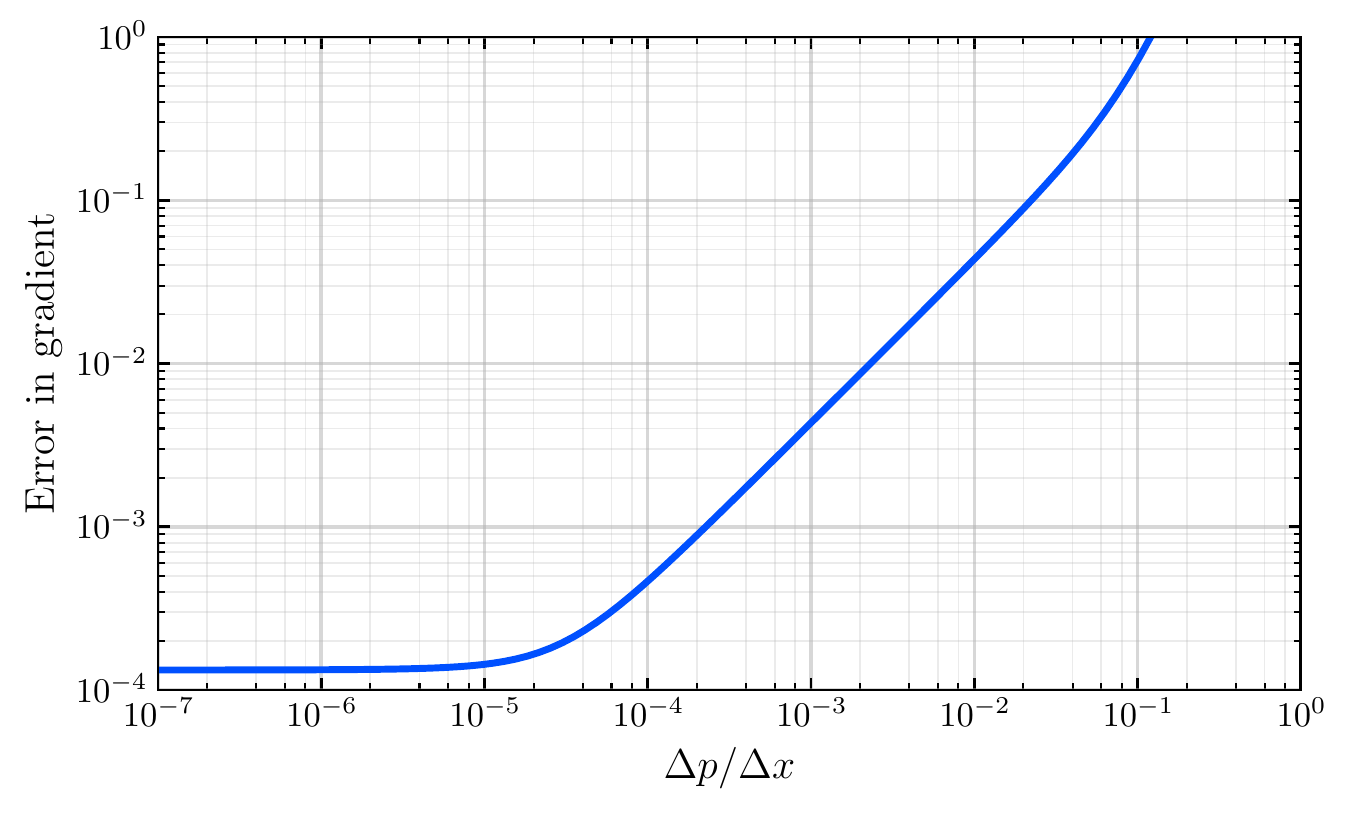}
		\caption{Accuracy of the gradient of the figure of merit computed using the adjoint method and grid smoothing as a function of the size of the perturbation $\Delta p$ used in the calculation (normalized to the width of a grid cell $\Delta x$).  The error in this gradient is defined with respect to the ``brute force'' gradient which is computed by sequentially perturbing each design variable of the system by a very small amount, running a new simulation, and then computing the new figure of merit.  Greater error is incurred as grid cells further from the original boundary of the shape contribute to the change in permittivity.  In order to achieve better than $\sim 1\%$ error, the step size used in the computation of the derivative $\partial A / \partial p_i$ has to be less than $\sim0.2\%$ of the grid spacing.} 
\label{fig:gradient_accuracy}
\end{figure}

Having defined a figure of merit and initial design, we are able to evaluate the accuracy of the gradients computed using the adjoint method with grid smoothing.  We do this by computing the gradient of $F$ using Equation (\ref{eq:dFdpi_final_simple}) and then comparing it to the brute-force calculation of the gradient.  The error between the gradient calculated using the adjoint method and the gradient calculated using brute-force finite differences is determined by evaluating

\begin{equation} \text{Error in } \mathbf{\nabla} F =
		\frac{\left|\mathbf{\nabla} F_\text{\tiny FD} - \mathbf{\nabla} 
		F_\text{\tiny AM}\right|}{\left|\mathbf{\nabla} F_\text{\tiny FD}\right|} 
		\label{eq:gradient_error} 
\end{equation}

\noindent where the subscripts ``FD'' and ``AM'' refer to ``finite difference'' and ``adjoint method,'' respectively.  The result of this comparison is shown in Fig. \ref{fig:gradient_accuracy} in which gradient accuracy is plotted against the size of the perturbation to each design variable ($\Delta p$) that is used to compute $\partial A / \partial p_i$.

When the perturbation to each design variable is sufficiently small (i.e. $\lesssim 10^{-5} \Delta x$ where $\Delta x$ is the grid spacing), the error in the gradient is much less than 1\% and is effectively independent of the size of the perturbation.  We attribute the non-zero minimum error to the presence of numerical error that arises during the forward and adjoint solution processes.  As the perturbation is made larger, the number of grid cells which contribute to the estimation of $\partial A / \partial p_i$ grows as the perturbed boundary intersects a greater and greater number of grid cells compared to the unperturbed boundary.  This in turn introduces additional error into the gradient calculation and is responsible for the nearly piecewise linear nature of the plotted gradient accuracy.  This result highlights the importance of having a truly continuous grid smoothing method.  Grid smoothing based on supersampling techniques is likely infeasible since the permittivity in each grid cell would have to be sampled tens or hundreds of thousands of times in order to achieve a high level of accuracy, a process which would be exceedingly computationally expensive.

Based on these results we choose the perturbation size to be $\Delta p = 10^{-7}\Delta x$ in order to ensure that the computed gradients are as accurate as possible.  These gradients are used with the Broyden-Fletcher-Goldfarb-Shanno (BFGS) algorithm, which we found worked best for this particular problem, in order to minimize the figure of merit and thus optimize the geometry of the waveguide taper.  

\begin{figure}[h]
    \centering
    \includegraphics[width=0.9\textwidth]{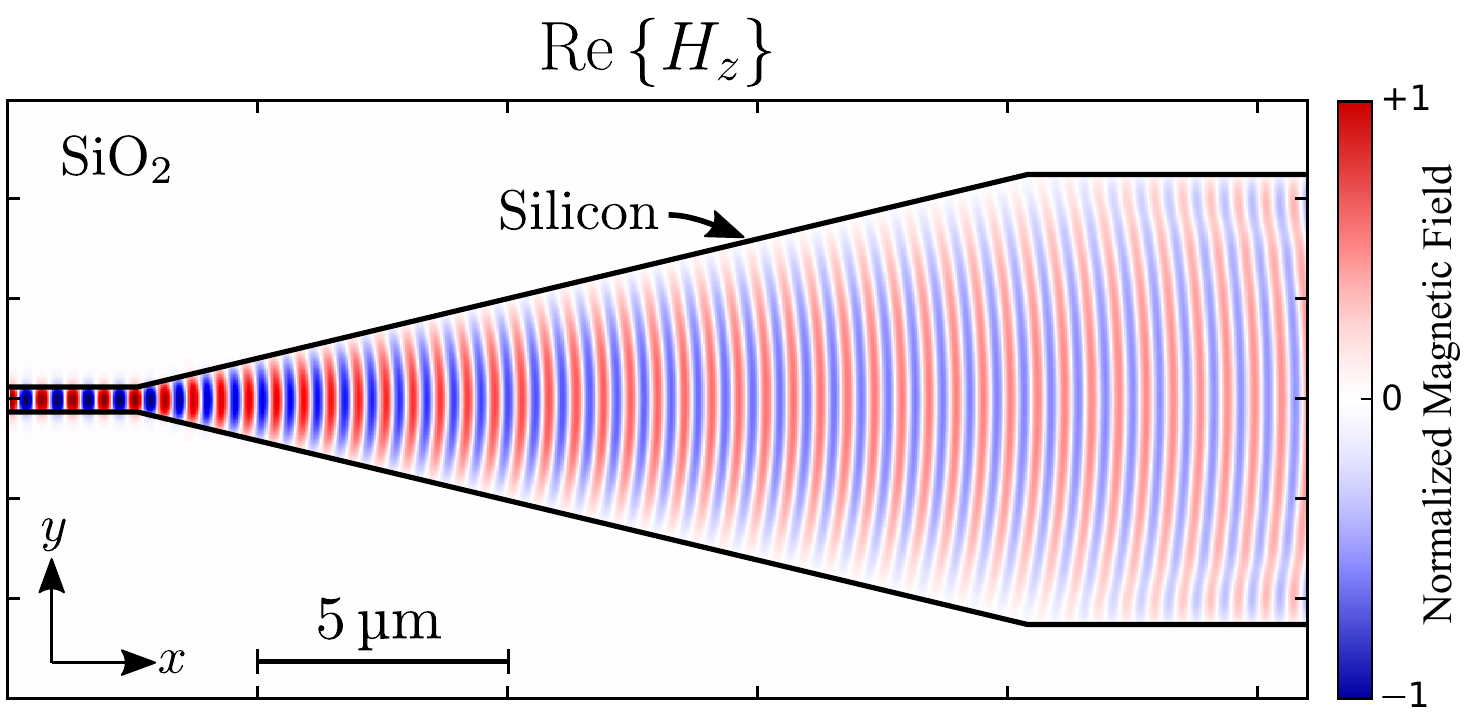}
		\caption{The initial structure overlaid with the real part of z-component of the magnetic field, $H_z$. Notice that the abrupt transition from the input to output waveguide results in significant curvature of the propagating wavefronts.  This results in significant coupling to higher order modes in the output waveguide, and the amount of power propagating in the fundamental mode of the output waveguide is only about 51\%.}
\label{fig:initial_Hz} 
\end{figure}

\begin{figure}[htp!] \centering
		\includegraphics[width=0.9\textwidth]{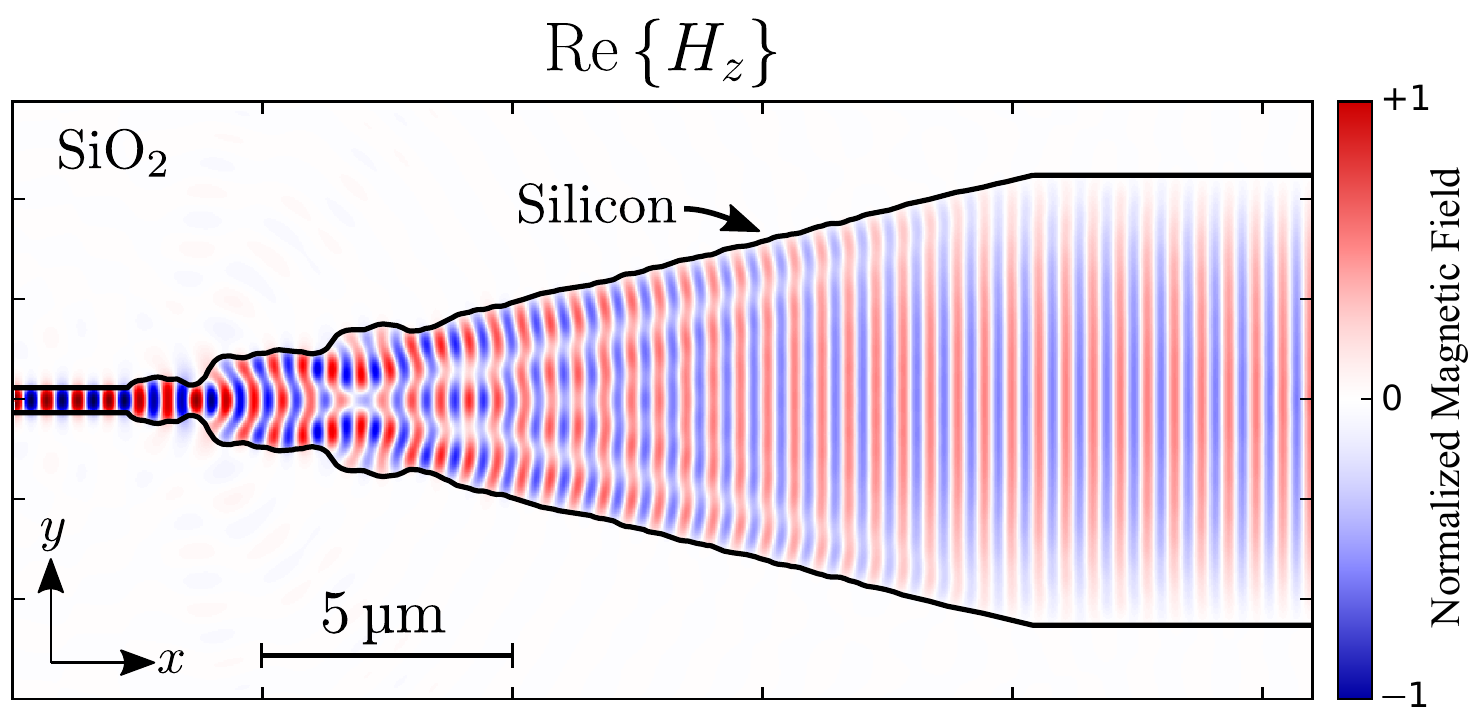}
		\caption{The final optimized structure overlaid with the real part of $H_z$. The optimized structure is very successful in coupling the input power into the fundamental mode of the output waveguide, and no curving of the outgoing wavefronts is visible.  This is reflected in the final efficiency of the structure of over 99\% at the design wavelength.} 
\label{fig:final_Hz} 
\end{figure}

The real part of $H_z$ is plotted for the initial design in Fig. \ref{fig:initial_Hz}.  As is clearly visible, the wavefronts become curved as they propagate along the linear taper.  This curving of the wavefronts results in significant coupling to higher order modes in the output waveguide, which is reflected by an initial efficiency of only 51\%. Beginning with this structure, we run the optimization until the figure of merit changes by less than $10^{-4}$ which takes 97 iterations of the minimization algorithm (which required 142 figure of merit and gradient calculations and $\sim$284 simulations in total).  The structure resulting from this optimization process is shown in Fig. \ref{fig:final_Hz}.  As desired, the wavefronts leaving the device are flat and there is no visible wave interference in the output waveguide that results from the excitation of higher order modes.  This final structure has an efficiency of just over 99\% (-0.041 dB) at the design wavelength of \SI{1550}{\nano\meter}.

\begin{figure}[h] \centering
		\includegraphics[width=1.0\textwidth]{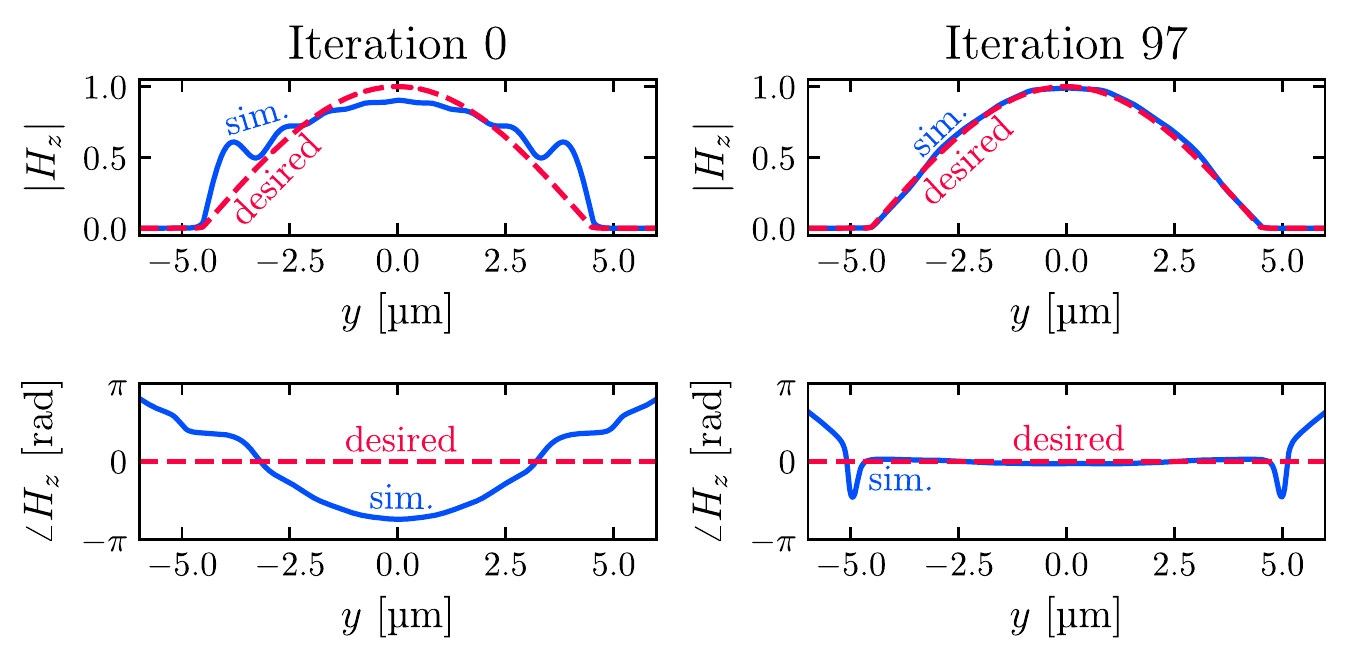}
		\caption{Slices of the fields shown in Fig. \ref{fig:initial_Hz} and Fig. \ref{fig:final_Hz} taken along a vertical line at the right edge of the simulation domain (where the figure of merit is computed).  The top row shows the magnitude of $H_z$ for the field at the beginning of the first iteration (left) and the field after the 97th iteration (right). The bottom row, meanwhile, shows the phase of $H_z$ for the first (left) and 97th (right) iteration.  The fields of the initial structure deviate significantly from the desired fields both in terms of amplitude and phase.  The fields of the optimize solution, meanwhile, match very closely to the desired fundamental mode, with its phase deviating only at the edges where the amplitude of the field is exceptionally low (and thus incurring minimal loss).} 
\label{fig:field_slices} 
\end{figure}

The full extent of the improvement in performance due to optimization is made clearly evident by comparing the slices of the magnetic field ($H_z$) taken perpendicular to the output waveguide for the first and last iteration as shown in Fig. \ref{fig:field_slices}.  In the first iteration, the simulated field amplitude and phase deviate significantly from the desired profiles.  In particular, the phase varies by approximately $\pi$ radians across the width of the output waveguide, which reflects the significant curvature visible in the full two dimensional plot of the fields and is indicative of the excitation of higher order modes.  

After the 97th iteration, on the other hand, the simulated magnetic field amplitude is almost indistinguishable from the desired amplitude profile.  Similarly, within the output waveguide, the simulated phase matches the desired phase, deviating by less than one tenth of a radian from the desired flat phase. While the phase deviates significantly outside of the waveguide, this is not a problem since the amplitude of the field is essentially zero in this region.  The close match in amplitude and phase to the desired field reflects the final calculated efficiency of just over 99\% at \SI{1550}{\nano\meter}.

\begin{figure}[htp!] \centering  
		\begin{subfigure}[c]{0.5\textwidth}
				\includegraphics[width=\textwidth]{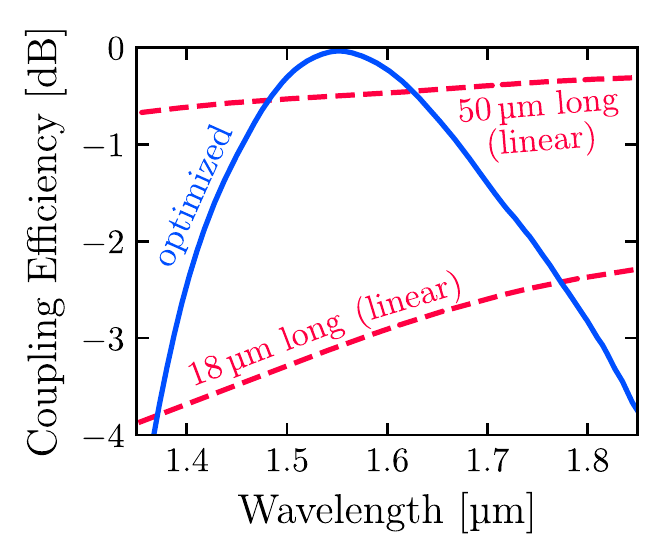}
				\caption{} 
				\label{fig:fom_bb}
		\end{subfigure} 
		\begin{subfigure}[c]{0.48\textwidth}
				\includegraphics[width=\textwidth]{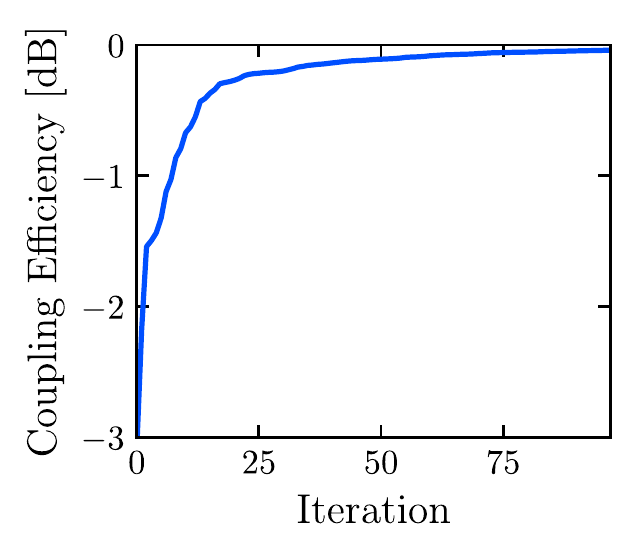}
				\caption{} 
				\label{fig:fom_iter}
		\end{subfigure}
		\caption{(a) Plots of  the final figure of merit versus wavelength and (b) the figure of merit versus iteration of the optimization.  The optimization achieves a figure of merit of over 90\% (\SI{-0.46}{\decibel}) in under 20 iterations, eventually reaching a final efficiency of just over 99\% (\SI{-0.041}{\decibel}) at the design wavelength. Although the optimization is only performed for a single wavelength, the final structure shows excellent broadband performance with a \SI{-3}{\decibel} bandwidth of approximately \SI{420}{\nano\meter}.  The final structure can thus be used in conjunction with other components such as grating couplers without significantly reducing the overall bandwidth of the combined component.}
				\label{fig:fom_iter_and_broadband} 
\end{figure}

In addition to the optimized design exhibiting extremely high efficiencies at the design wavelength, it also performs favorably over a wide range of wavelengths as demonstrated by Fig. \ref{fig:fom_bb}.  The optimized design exhibits a \SI{3}{\decibel} bandwidth of approximately \SI{420}{\nano\meter}.  This wide bandwidth makes these tapers well suited for use with grating couplers whose bandwidth is generally less than $\sim$\SI{100}{\nano\meter}.  Because the efficiency of the taper does not drop by more than roughly one third of a \SI{}{\decibel} over a \SI{100}{\nano\meter} range about the design wavelength, the combined bandwidth of the optimized taper plus a typical silicon grating coupler would not be significantly reduced compared to the bandwidth of the grating coupler alone.

It is interesting to note that our optimized waveguide taper exhibits a stronger sensitivity compared to the original linear taper.  This is likely a result of the introduction of smaller wavelength-scale features by the optimization. This result is not surprising, however, because we considered only a single wavelength when computing our figure of merit. We partially mitigate this issue by enforcing a radius of curvature constraint; by limiting the minimum size of features, we avoid the development of very fine structures which tend to exhibit a stronger wavelength dependence.  

Despite the increase in sensitivity to wavelength, our optimized design still performs better than longer linear tapers over a limited bandwidth.  For example, the optimized taper has a higher efficiency than a \SI{50}{\micro\meter} linear taper over a \SI{144}{\nano\meter} range as shown in Fig. \ref{fig:fom_bb}.  Similarly, compared to a \SI{100}{\micro\meter} long linear taper, our optimized design has a higher efficiency over an \SI{80}{\nano\meter} bandwidth.  Thus in cases where a more modest bandwidth is needed, the optimized design can achieve higher efficiencies than can be achieved by linear tapers that are more than five times as long.  In order to achieve equal performance at the design wavelength of \SI{1550}{\nano\meter}, a linear taper with a length of over \SI{180}{\micro\meter} (an order of magnitude longer than the optimized design) is required.

The strength of the optimization methods we have employed not only lie in the performance of the final result, but also in the method's ability to rapidly converge to an optimal solution.  Consider how the efficiency of the taper evolves with each iteration of the optimization plotted in Fig. \ref{fig:fom_iter}.  Within the first two iterations of the optimization alone, the figure of merit is increased by approximately 20\% (from \SI{-3}{\decibel} to \SI{-1.5}{\decibel}).  After the first 20 iterations, the figure of merit has further increased to over 90\% (\SI{-0.46}{\decibel}), which is then followed by slower improvement as the gradients of the figure of merit become smaller and smaller.  Convergence to the final efficiency is smooth as a result of the minimization algorithm used.  

It is worth noting that the minimization algorithm we used is a local optimization method which makes no guarantees that a global optimum is found. In general, however, finding global optima would be ideal.  Heuristic techniques like the genetic algorithm are often employed in electromagnetics in order to increase the likelihood of finding a global optimum.  These algorithms often require thousands of simulations before an optimum is found \cite{johnson_genetic_1997,roelkens_high_2006}, with no guarantees that it is a global optimum.  The potential advantages are thus not always perfectly evident.  In many problems, however, we can largely eliminate these concerns by taking advantage of our own physical intuition of the problem and ``guessing'' a good starting design.  In the case of our waveguide taper, for example, we know that linear transitions (which work well at longer lengths) are well behaved, exhibit low reflections, and suffer primarily from diffraction effects that cause the propagating wavefront to become curved.  It is thus not surprising that beginning with the linear taper leads to a local optimum with such high efficiency and wide bandwidth.

\section{Conclusion}

This example demonstrates the great utility of shape optimization in electromagnetics: given a goal (i.e. figure of merit) and an initial guess for a design, we can quickly design passive components which outperform anything we could have otherwise designed by hand.  This is made possible by grid smoothing which gives us substantial freedom in how we represent material boundaries and enables us to calculate accurate gradients of a figure of merit with respect to large numbers of design parameters when used with the adjoint method.  With these accurate gradients at our disposal, we can employ a wide range of very powerful minimization algorithms.  Many of these minimization methods can be used in place of BFGS for electromagnetic shape optimization without any modification to this work.

Combining our physical intuition of electromagnetics with gradient-based shape optimization provides us with a path forward to improve many of the components in integrated optics.  Based on the success of topology and shape optimization in other fields and the ease with which it can be applied to electromagnetics, we expect inverse electromagnetic design to become an essential component of the future nanophotonic engineer's toolbox. To facilitate this, we have released our own optimization tool, of which grid smoothing is a core component, as open source software \cite{michaels_emopt_2018}.

\appendix \section{Mode Matching Integral} \label{sec:appendix_mode_match}

Mode overlap, or mode-matched efficiency, refers to the degree to which an incident field can couple into a desired mode of a system. Computing the mode overlap is essential to determining the efficiency of many optical devices and is therefore highly relevant to shape optimization. For the reader’s benefit, we present a relatively detailed derivation of the mode overlap integral here.

Our first step determining the mode-matched efficiency is to express our input field as a sum of the allowed propagating modes of the system.  Specifically, the basis we will use consists of the electric and magnetic fields of both forward and backward traveling waves. We begin by writing the electric field as a sum of these basis functions,

\begin{align} \mathbf{E} &= \mathbf{E}_\mathrm{fwd} + \mathbf{E}_\mathrm{back}
		\\ &= \sum\limits_m \left( a_m \mathrm{e}^{i k_m z} + b_m
		\mathrm{e}^{-i k_m z} \right) \mathbf{E}_m 
		\label{eq:E_field_expansion}
\end{align}

\noindent where the two terms in parentheses correspond to the forward and backward components and are given separately by

\begin{align} \mathbf{E}_\mathrm{fwd} &= \sum\limits_m a_m \mathrm{e}^{i k_m z}
		\mathbf{E}_m \label{eq:E_expansion_fwd} \\ \mathbf{E}_\mathrm{back} &=
		\sum\limits_m b_m \mathrm{e}^{-i k_m z} \mathbf{E}_m \;\; .
		\label{eq:E_expansion_back} 
\end{align}

\noindent In the expressions above, $\mathbf{E}_m$ is the electric field profile of the $m$th mode of the system. The magnetic field, meanwhile, can be written in a similar form.  Applying Faraday's Law and assuming harmonic time dependence of the electric field, the magnetic field can be written as an expansion of the forward and backward wave

\begin{align} \mathbf{H} &= \mathbf{H}_\mathrm{fwd} + \mathbf{H}_\mathrm{back}
		\\ &= \sum\limits_m \left( a_m \mathrm{e}^{i k_m z} - b_m
		\mathrm{e}^{-i k_m z} \right) \mathbf{H}_m \label{eq:H_field_expansion}
\end{align}

Notice in Equation (\ref{eq:H_field_expansion}), that the backward propagating term is preceded by a negative sign. This arises out of the requirement that power flow in the negative direction (and hence the Poynting vector, $\mathbf{E}_m \times \mathbf{H}_m$ point in the negative direction). In both Equation (\ref{eq:H_field_expansion}) and (\ref{eq:E_field_expansion}), we have chosen to write the fields as a sum of the forward and backward traveling waves. In general, given an arbitrary field, we will not know the forward and backward traveling components but only their sum.  Our goal now is to develop the machinery needed to separate the different forward and backward traveling modes which compose an arbitrary field.  

This is equivalent to finding the coefficients $a_m$ and $b_m$.  To do so, we must take advantage of the orthogonality condition of our electromagnetic basis which arises as a result of Lorentz reciprocity and is given by\cite{chen_appendix_2006}

\begin{equation} \frac{\displaystyle\iint\limits_A d\mathbf{A} \cdot
		\mathbf{E}_m \times \mathbf{H}_n^{*}}{\displaystyle\iint\limits_A
		d\mathbf{A} \cdot \mathbf{E}_m \times \mathbf{H}_m^{*}} = \delta_{m n} 
		\label{eq:orthogonality} 
\end{equation}

\noindent where $\delta_{m n}$ is the Kronecker delta.  We apply this orthogonality condition by computing the surface integral of $\mathbf{E} \times \mathbf{H}_m^*$ and rearranging terms yields an expression relating $a_m$ and $b_m$ to the electric field and $m$th mode

\begin{equation} a_m \mathrm{e}^{i k_m z} + b_m \mathrm{e}^{-i k_m z} =
		\frac{\iint\limits_A d\mathbf{A} \cdot \mathbf{E} \times \mathbf{H}_m^{*}}{S_m} 
		\label{eq:coeff_1} 
\end{equation}

\noindent where $S_m$ is related to the power propagating in the $m$th mode

\begin{equation} S_m = \displaystyle\iint\limits_A d\mathbf{A} \cdot
		\mathbf{E}_m \times \mathbf{H_m}^* \;\; .  
		\label{eq:S_m}
\end{equation}

\noindent A second expression for $a_m$ and $b_m$ can be found by computing the surface integral of $\mathbf{E}_m \times \mathbf{H}^*$ which yields

\begin{equation} a_m \mathrm{e}^{i k_m z} - b_m \mathrm{e}^{-i k_m z} =
		\frac{\iint\limits_A d\mathbf{A} \cdot \mathbf{E}_m^* \times
		\mathbf{H}}{S_m^*} \;\; .  \label{eq:coeff_2} \end{equation}

With two equations and two unknowns, we are now able to solve for the coefficients.  Adding the two equations produces an expression for $a_m$

\begin{align} 
		a_m = \frac{1}{2} \mathrm{e}^{-i k_m z} \left(
		\frac{\iint\limits_A d\mathbf{A} \cdot \mathbf{E} \times
		\mathbf{H}_m^{*}}{S_m} + \frac{\iint\limits_A d\mathbf{A} \cdot
		\mathbf{E}_m^* \times \mathbf{H}}{S_m^*} \right) \label{eq:a_m} 
\end{align}

\noindent  while subtracting them yields and expression for $b_m$. Using these equations, decomposing a given field into the modes of the system is a straightforward calculation.  Mode matching requires that we take this one step further and determine how much \emph{power} is propagating in a desired mode.  To find this, we compute the power propagating through a plane in the forward wave, i.e. $P_\mathrm{fwd} = \frac{1}{2} \Re\{ \iint d\mathbf{A} \cdot \mathbf{E}_\mathrm{fwd} \times \mathbf{H}_\mathrm{fwd}^* \}$.  This calculation results in a sum whose terms correspond to the power propagating in each mode.  This power is more conveniently expressed as a fraction of the total power propagating in the field:

\begin{equation} \eta_m^{\mathrm{fwd}} = \frac{P_m}{P_\mathrm{in}} = |a_m|^2
		\frac{\Re\left\{ S_m \right\} }{ \Re\left\{  \iint\limits_A d\mathbf{A}
		\cdot \mathbf{E} \times \mathbf{H}^* \right\} } \;\; .
		\label{eq:mode_overlap} 
\end{equation}

\noindent Equation (\ref{eq:mode_overlap}) describes the amount of power that can couple from an incident field into a desired mode of the system.  This expression can be further simplified by noticing that in most problems, a backward traveling wave is not present at the output of the system. In this case, $b_m$ equals zero and as a result

\begin{equation} \frac{\iint\limits_A d\mathbf{A} \cdot \mathbf{E} \times
		\mathbf{H}_m^{*}}{S_m} = \frac{\iint\limits_A d\mathbf{A} \cdot
		\mathbf{E}_m^* \times \mathbf{H}}{S_m^*} \label{eq:b_m_is_0}
\end{equation}

\noindent Taking this into account allows us to simplify Equation (\ref{eq:mode_overlap}) to  

\begin{equation} 
	\eta_m =  \frac{ \frac{1}{2}\Re \left\{\iint\limits_A
	d\mathbf{A} \cdot \mathbf{E}_m \times \mathbf{H}_m^*\right\}}{\frac{1}{2}\Re \left\{\iint\limits_A
	d\mathbf{A} \cdot \mathbf{E} \times \mathbf{H}^*\right\}}\frac{\left|\iint\limits_A
	d\mathbf{A} \cdot \mathbf{E} \times \mathbf{H}_m^*\right|^2}{\left|\iint\limits_A
	d\mathbf{A} \cdot \mathbf{E}_m \times \mathbf{H}_m^*\right|^2}
	\label{eq:coupling_efficiency_almost_there}
\end{equation}

\noindent where $\mathbf{E}_m$ and $\mathbf{H}_m$ are the field profiles that we desire the grating to generate.  This equation is the most general form of the mode-matched efficiency.  However, in many applications, $\mathbf{E}_m$ and $\mathbf{H}_m$ will correspond to a guided mode.  In this case, we can further simplify the expression by noting that for a guided or free space mode, the integral $\iint\limits_A d\mathbf{A} \cdot \mathbf{E}_m \times \mathbf{H}_m^*$ is real valued.  In this case we can cancel the first term in the numerator and we find that the mode overlap is given by

\begin{equation} \eta_{m,\mathrm{guided}} =  \frac{1}{4 P_m P_\mathrm{in}}
		\left|\iint\limits_A d\mathbf{A} \cdot \mathbf{E} \times
		\mathbf{H}_m^*\right|^2 \label{eq:coupling_efficiency_simplified}
\end{equation}

\noindent where we have chosen to write 

\begin{align*} 
    P_m &= \frac{1}{2}\Re \left\{\iint\limits_A d\mathbf{A} \cdot \mathbf{E}_m
		\times \mathbf{H}_m^*\right\} \\ 
        P_\mathrm{in} &= \frac{1}{2}\Re
		\left\{\iint\limits_A d\mathbf{A} \cdot \mathbf{E} \times
		\mathbf{H}^*\right\} 
\end{align*}

\noindent which describe the power in the incident field and the power in the desired mode (neither of which are guaranteed to be normalized to unity power). In many situations, we wish to know the total efficiency with which a device will output into a desired mode.  This efficiency is differs slightly from the mode overlap expression given above as it compares the fraction of power coupled into a desired mode at the output of a device to the \emph{total} power input to the device.  This minor difference is easily accounted for by replacing $P_\mathrm{in}$ with $P_\mathrm{src}$, the total source power of the system.


\bibliographystyle{ieeetr} \footnotesize\bibliography{references} 

\end{document}